\documentclass[a4paper,journal,twoside]{IEEEtran}
\IEEEoverridecommandlockouts
\newcommand\hmmax{0}
\newcommand\bmmax{0}

\usepackage{graphicx}
\usepackage{tikz}
\usepackage{pgfplots}
\usepackage{xcolor}
\usepackage{color, colortbl}
\usepackage{amsmath}
\usepackage{bbm}
\usepackage{amsfonts, amssymb}
\usepackage{mathtools}
\usepackage{cite}
\usepackage[nolist]{acronym}
\usepackage{amsmath}

\usepackage{booktabs} 		% cute tables
\usepackage{diagbox}		% diagonal cell division in table
\usepackage{upgreek}
\usepackage{bbm}
\usepackage{Files/bm}
\usepackage{Files/winsnotation}
\usepackage{Files/Symbols_OPL}
\usepackage{Files/LVcolours} % cute colours

\usepackage[ruled,vlined,noresetcount]{algorithm2e}
\usepackage{setspace}	 	% stretch on algorithm2e
\usepackage{comment}
\usepackage{physics}

\usepackage{array}
\newcolumntype{C}[1]{>{\centering\arraybackslash}p{#1}}

\makeatletter
\def\endthebibliography{%
  \def\@noitemerr{\@latex@warning{Empty `thebibliography' environment}}%
  \endlist
}
\makeatother
\usepackage{amsthm}
\theoremstyle{definition}

\newcommand{\px}{p_{\mathrm{X}}}
\newcommand{\py}{p_{\mathrm{Y}}}
\newcommand{\pz}{p_{\mathrm{Z}}}

\newcommand{\PauliX}{\M{X}}
\newcommand{\PauliY}{\M{Y}}
\newcommand{\PauliZ}{\M{Z}}

\pgfplotsset{compat=1.17}
\begin{document}

%\title{Performance Analysis of Quantum \\ Topological Planar Code}
\title{Logical Error Rates of XZZX and Rotated Quantum Surface Codes}
%\title{Performance Analysis of Quantum \\ Error-Correcting Surface Codes}

%\author{%
%  \IEEEauthorblockN{tbd}
%  \IEEEauthorblockA{CNIT/WiLab, DEI, University of Bologna, Italy\\
%  	Email: \{ \}@unibo.it }
%}

%\author{%
%    \IEEEauthorblockN{Diego Forlivesi, Lorenzo Valentini, Marco Chiani \\}
%    \IEEEauthorblockA{CNIT/WiLab, DEI, University of Bologna, Italy \\
%  	Email: \{diego.forlivesi2, lorenzo.valentini13, marco.chiani\}@unibo.it }
%}

\author{Diego Forlivesi,~\IEEEmembership{Student~Member,~IEEE,}
Lorenzo~Valentini,~\IEEEmembership{Student~Member,~IEEE,}
        and~Marco~Chiani,~\IEEEmembership{Fellow,~IEEE}
\thanks{The authors are with the Department of Electrical, Electronic, and Information Engineering ``Guglielmo Marconi'' and CNIT/WiLab, University of Bologna, 40136 Bologna, Italy. E-mail: \{diego.forlivesi2,lorenzo.valentini13, marco.chiani\}@unibo.it. 
%\thanks{
Work supported by Ministero dell’Università e della Ricerca, PNRR project PRIN n. 2022JES5S2. 
%Part of this work will be presented at IEEE ICC 2023, Rome, May 2023 ({\it arXiv:2302.13015}).
}
}

% make the title area
% Don't write page number 0 to the cover page.
\maketitle 
%\markboth{Submitted to IEEE JSAC}{}

\begin{acronym}
% usage: \ac{SW}, \acp{SW} for plurals \acf{SW} Use the full name of the acronym.
%\acs{SW}Use the acronym, even before the first corresponding \ac command
%\acl{acronym}Expand the acronym without using the acronym itself.
\small
\acro{AWGN}{additive white Gaussian noise}
\acro{BCH}{Bose–Chaudhuri–Hocquenghem}
\acro{CDF}{cumulative distribution function}
\acro{CRC}{cyclic redundancy code}
\acro{LDPC}{low-density parity-check}
\acro{ML}{maximum likelihood}
\acro{MWPM}{minimum weight perfect matching}
\acro{QECC}{quantum error correcting code}
\acro{PDF}{probability density function}
\acro{PMF}{probability mass function}
\acro{MPS}{matrix product state}
\acro{WEP}{weight enumerator polynomial}
\acro{WE}{weight enumerator}
\acro{BD}{bounded distance}
\acro{QLDPC}{quantum low density parity check}

\end{acronym}
\setcounter{page}{1}

\begin{abstract}
% An important class of quantum error-correcting codes is constituted by surface codes, which are stabilizer codes with a planar geometry that facilitates implementations. The original proposal, where the generators are composed of only Pauli $\PauliX$ operators or only Pauli $\PauliZ$ operators in a square structure, can be modified by rotating the lattice, and also by mixing the generators in the so-called XZZX variant. Despite their importance, a theoretical analysis of the logical error rate for XZZX and rotated surface codes has not been investigated. 
% In this paper, we provide theoretical formulas for these variants of surface codes, based on recent results on the weights distribution of stabilizer codes. 
% For example, on an asymmetric channel with asymmetry $A=10$ and physical error rate $p \to 0$, it is found that the logical error rate $p_\mathrm{L}$ is asymptotically $p_\mathrm{L} \to 10 p^2$ for the rotated $[[9,1,3]]$ XZZX code, and $p_\mathrm{L} \to 18.3 p^2$ for the $[[13,1,3]]$ surface code.
% We also found, for asymmetric channels, that if we start from a rectangular lattice, we should not apply both the rotation and the XZZX variants.

Surface codes are versatile quantum error-correcting codes known for their planar geometry, making them ideal for practical implementations. While the original proposal used Pauli $\PauliX$ or Pauli $\PauliZ$ operators in a square structure, these codes can be improved by rotating the lattice or incorporating a mix of generators in the XZZX variant. However, a comprehensive theoretical analysis of the logical error rate for these variants has been lacking. 
To address this gap, we present theoretical formulas based on recent advancements in understanding the weight distribution of stabilizer codes. For example, over an asymmetric channel with asymmetry $A=10$ and a physical error rate $p \to 0$, we observe that the logical error rate asymptotically approaches $p_\mathrm{L} \to 10 p^2$ for the rotated $[[9,1,3]]$ XZZX code and $p_\mathrm{L} \to 18.3 p^2$ for the $[[13,1,3]]$ surface code. 
Additionally, we observe a particular behavior regarding rectangular lattices in the presence of asymmetric channels. Our findings demonstrate that implementing both rotation and XZZX modifications simultaneously can lead to suboptimal performance. Thus, in scenarios involving a rectangular lattice, it is advisable to avoid using both modifications simultaneously. 
This research enhances our theoretical understanding of the logical error rates for XZZX and rotated surface codes, providing valuable insights into their performance under different conditions. %These findings empower informed design decisions in the realm of quantum error correction applications.

\end{abstract}


%%%%%%%%%%%%%%%%%%%%%%%%%%%%%%%%%%%%%%%%%%%%%%%%%%%%%%
\section{Introduction}
%%%%%%%%%%%%%%%%%%%%%%%%%%%%%%%%%%%%%%%%%%%%%%%%%%%%%%
%IBM Extracting the full potential of computation and realizing quantum algorithms with a super-polynomial speedup will most likely require major advances in quantum error correction technology. 
%IBM Quantum error correcting codes provide a similar redundant representation of quantum states that protects them from certain types of errors. 
The Quantum Internet has emerged as a promising paradigm with the potential to revolutionize various fields, including computation, secure communication, and enhanced sensing. For example, it will enable the realization of distributed quantum computers which could solve complex problems more efficiently than classical computers. However, quantum systems are highly susceptible to errors and decoherence, which pose significant challenges to realizing reliable quantum computation and quantum networks \cite{Sho:95,NieChu:10,Kim:08,MurLiKim:16,WehElkHan:18,CacCalVan:20,Val23:PurifCoding}. Quantum error-correcting codes play a crucial role in preserving and protecting quantum information and are therefore fundamental in quantum computation, quantum memories, and quantum communication systems \cite{Sho:95,NieChu:10,Got:09,Ter:15,Bab:19,ChiConWin:20,OstOrsLaz:22,ZorDePGio:23}.
%\cite{Sho:95,Laf:96,Ste:96a,Got:96,Kni:97,FleShoWin:08,Got:09,NieChu:10,Ter:15,MurLiKim:16,Rof:19,Bab:19,ChiConWin:20,RehShi:21,ZorDePGio:23}.

Among the different types of \acp{QECC}, surface codes have gained significant attention, as they are compatible with 2D architectures, and also require error syndrome measurements which are inherently among few close qubits. The original surface codes, which pertain to the class of stabilizer codes, adopt generators composed of three or four Pauli operators, each generator using one Pauli type only \cite{BraKit:98,FowMarMar:12}. Variants of the original surface codes have been proposed. In the so-called ``rotated'' surface codes, some qubits are removed, so that the generators have only two or four Pauli operators, maintaining the same minimum distance \cite{HorFowDev:12}.  Another variant consists in allowing to mix Pauli type operators within each generator, in the so-called XZZX variant~\cite{AtaTucBar:21}.

Implementations of surface codes with minimum distance three have been reported recently in \cite{KriSebLac:22} and \cite{ZhaYouYe:22}. In these works, 17 physical qubits (9 data qubits and 8 qubits for syndrome measurements) in a superconducting circuit have been used to realize a rotated squared surface code, employing stabilizer generators of type XX, ZZ, XXXX, ZZZZ. The error correction is performed with a \ac{MWPM} algorithm and is repeated to preserve one logical qubit from decoherence. 
In \cite{AchRajAle:22} a larger surface code is realized, with superconducting devices used to implement a 49 physical qubits code (25 data qubits and 24 qubits for syndrome measurements) with minimum distance five. This code slightly outperforms the distance three surface code, demonstrating a possible path toward scalable quantum error correction. The architecture adopted in \cite{AchRajAle:22} is a rotated squared surface code with XX, ZZ, and XZZX generators.  

The theoretical performance of surface codes has been investigated in the literature mainly in terms of accuracy threshold over symmetric channels for fault-tolerant quantum computing \cite{BraKit:98,FowMarMar:12,BraSerSuc:14,AtaTucBar:21}. 
In \cite{Ash00:MacwPartI, Ash00:MacwPartII}, an analysis using the MacWilliams identities was proposed to derive theoretical bounds for quantum error detection. 
Recently, a new methodology has been introduced to analyze the performance of generic stabilizer codes, starting from the MacWilliams identities to derive the logical errors weight distribution \cite{ForValChi:23}. 
The method allows to derive closed-form formulas for the logical error rate vs. physical error rate, for both the depolarizing channel, as well as  for asymmetric channels. These are channels  where errors described by Pauli X, Z, and Y operators may have different occurrence probabilities \cite{FleShoWin:08,SarKlaRot:09,ChiVal:20a,RehShi:21}. 

This article aims to provide a theoretical analysis of the logical error rate for rotated and XZZX quantum surface codes, which are the most investigated in practical implementations \cite{KriSebLac:22,ZhaYouYe:22,AchRajAle:22}. 
The key contributions of the paper can be summarized as follows:
\begin{itemize}
    \item we derive the \ac{WE} for the undetectable errors of rotated and/or XZZX surface codes, via MacWilliams identities;
    \item we provide a set of coefficients that can be used to obtain the logical error probability without implementing the decoder, for the main surface codes. This feature can be used at higher level in the protocol stack to aid the design of Quantum Internet;
    %\item we provide tables of coefficients which permit to express in closed form a tight approximation of the logical error rate under \ac{MWPM} decoding, over arbitrary asymmetric quantum channels;
    \item we compare the original surface codes with the rotated and/or XZZX variants, highlighting the differences and advantages over the depolarizing channel and over asymmetric channels;
    \item we show that the combined use of rotation and XZZX variants in rectangular lattice leads to a degradation in performance.
\end{itemize}
%
%Since the current technology allows few qubit architectures, numerical results are shown for the shortest topological codes to prove the effectiveness of the proposed approach.    
 We will focus in particular on the $[[9,1,3]]$, the $[[25,1,5]]$, and the  $[[15,1,3/5]]$ rotated surface codes, with and w/o the XZZX variant, and compare them with the original $[[13,1,3]]$, the  $[[41,1,5]]$, and the $[[23,1,3/5]]$ surface codes. In this work, we have included the shortest examples for both symmetric and asymmetric variants of surface codes. This decision is driven by the current limitations in quantum computing, which only supports a restricted number of physical qubits. Nonetheless, our analytical approach is general and, in principle, can be applied to derive the logical error rates of arbitrarily large surface codes.  %, with the \acf{MPS} decoder and the \acf{MWPM} decoder. 

This paper is organized as follows. Section~\ref{sec:preliminary} introduces preliminary concepts and models together with some background material. In Section~\ref{sec:quantum_top} we analyze rotated, XZZX, and  rotated XZZX surface codes, obtaining the \ac{WE} for the undetectable errors from MacWilliams identities and applying it to the evaluation of the logical error rate. 
In Section~\ref{sec:NumRes} we provide the error rate comparison and discuss the effect of rotation on rectangular surface lattices.  %Numerical results are discussed in Section~\ref{sec:NumRes}. %Finally, conclusions are drawn in Section~\ref{sec:conclusions}.

%%%%%%%%%%%%%%%%%%%%%%%%%%%%%%%%%%%%%%%%%%%%%%%%%%%%%%
\section{Preliminaries and Background}
\label{sec:preliminary}

\subsection{Quantum Stabilizer Error-Correcting Codes}
\label{subsec:QEC}

We indicate with $[[n,k,d]]$ a \ac{QECC} with minimum distance $d$, that encodes $k$ logical qubits  $\ket{\varphi}$ into a codeword of $n$ data qubits  $\ket{\psi}$, allowing the decoder to correct all patterns up to $t = \lfloor(d-1)/2 \rfloor$ errors. A complete decoder will also correct some patterns of more than $t$ errors.  
The Pauli operators are indicated as $\PauliX, \PauliY$ and $\PauliZ$. Using the stabilizer formalism, each code is represented by $n-k$ independent and commuting operators $\M{G}_i \in \mathcal{G}_n$, called stabilizer generators (or simply generators), where $\mathcal{G}_n$ is the Pauli group on $n$ qubits \cite{Got:09,NieChu:10}.   
The subgroup of $\mathcal{G}_n$ generated by all combinations of the $\M{G}_i \in \mathcal{G}_n$ is a stabilizer, indicated as $\mathcal{S}$.  
The code $\mathcal{C}$ is the set of quantum states $\ket{\psi}$ stabilized by $\mathcal{S}$, i.e., satisfying 
$\M{S}\ket{\psi}=\ket{\psi}, \, \forall \M{S} \in \mathcal{S}$, or, equivalently, 
$\M{G}_i \ket{\psi}=\ket{\psi},\, i=1, 2, \ldots, n-k$. 
The generators have great importance in quantum error correction since they describe which measurements on the quantum codewords do not perturb the original quantum state. This enables the possibility to perform quantum error correction exploiting error syndrome decoding \cite{NieChu:10}.
The logical operators of a stabilizer code are those that commute with the stabilizer group but are not contained in it.
Hence, these operators map a codeword into a different one. %: if the error introduced by a quantum channel is a logical operator, the error cannot be detected. 
The knowledge of the logical operators' structure permits the evaluation of the channel errors that, together with the decoding correction, cause a decoding failure. As shown in \cite{ForValChi:23}, the logical error rate thus depends on the stabilizer logical operators \ac{WE} and on the employed decoder (e.g., based on the \ac{MWPM} algorithm).
%This approach allows us to analytically derive. 
We remark that the actual logical error rates attainable by a complete decoder, like the \ac{MWPM}, are beyond the conventional bounded distance performance.

%%%%%%%%%%%%%%%%%%%%%%%%%%%%%%%%%%%%%%%%%%%%%%%%%%%%%%

\subsection{Theoretical Performance Analysis of Quantum Codes}
We analyze codes over quantum channels assuming the different qubits in the code experience independent and identically distributed errors, a model which is commonly adopted to compare quantum error correcting codes \cite{AtaTucBar:21, SarKlaRot:09, FowSteGro:09, MacMitMcF:04, BraCroGam:23, CacCalVan:20, NieChu:10, OstOrsLaz:22}.

Specifically, we assume a quantum channel characterized by the errors $\M{X}$, $\M{Z}$ or $\M{Y}$ occurring with probabilities $\px$, $\pz$, and $\py$, respectively. The probability of a generic error on a qubit is $p = \px + \pz + \py$.

Among these quantum channels, two important models are the \emph{depolarizing channel} where $\px = \pz = \py = p / 3$, and the \emph{phase-flip channel} where $p=\pz$, $\px = \py=0$. 
We characterize an asymmetric channel by the asymmetry parameter $A = 2\pz /(p - \pz)$. 
In this way, varying the asymmetry $A$ we can evaluate the performance of channels having a prevalent error type (i.e., $\M{Z}$), passing from the depolarizing channel ($A = 1$) up to the phase-flip channel ($A \to \infty$).
%, together with channels having a dominance of one error type (i.e., $\M{Z}$).
Error-correcting codes can also be designed to exhibit asymmetric behaviors with respect to $\M{X}$, $\M{Y}$ and $\M{Z}$ Pauli errors \cite{SarKlaRot:09,ChiVal:20a}. We adopt the notation $[[n, k,d_\mathrm{X}/d_\mathrm{Z}]]$ for asymmetric codes able to correct all patterns up to $t_\mathrm{X} = \lfloor(d_\mathrm{X}-1)/2\rfloor$ Pauli $\M{X}$ errors and $t_\mathrm{Z} = \lfloor(d_\mathrm{Z}-1)/2\rfloor$ Pauli $\M{Z}$ errors. 

To evaluate the performance beyond the nominal error correction capability of the code, we define $f_j(i,\ell)$ as the fraction of errors of weight $j$, with $i$ Pauli $\M{Z}$ and $\ell$ Pauli $\M{X}$ errors, which are not corrected by a given decoder. Hence, the error probability of a \ac{QECC}, also indicated as logical error rate, can be written as \cite{ForValChi:23}
\begin{align}
\label{eq:rho_L}
    p_L = \sum_{j = 0}^{n} \binom{n}{j}(1-p)^{n-j} p^j (1-\beta_j)
\end{align}
where
\begin{align}
\label{eq:betaGen}
    \beta_j =1- \frac{1}{p^j}\sum_{i = 0}^{j}\binom{j}{i} \, \pz^i \, \sum_{\ell = 0}^{j-i} \binom{j-i}{\ell}\, \px^\ell \, \py^{j-i-\ell} f_j(i,\ell)\,
\end{align}
is the probability that an error of weight $j$ is correctly decoded. 
Note that $f_j(i,\ell)$ depends only on the particular decoder implementation, while $\beta_j$ also depends on the quantum channel parameters.
In the following, we will consider channels with asymmetry $A$, so that  \eqref{eq:betaGen} can be rewritten 
\begin{align}
\label{eq:betaA}
    \beta_j=\beta_j(A) =1- \frac{1}{(A+2)^j}\sum_{i = 0}^{j}\, A^i\,\sum_{\ell = 0}^{j-i} \binom{j}{i} \binom{j-i}{\ell} f_j(i,\ell)\,.
\end{align}

For $p \ll 1$ and symmetric codes, the logical error rate is well-approximated by 
\begin{align}
\label{eq:RhoLSymApprox}
p_\mathrm{L} 
&\approx \left(1-\beta_{t+1}\right) \binom{n}{t+1}p^{t+1} %\,.%+ (1-\beta_{t+2}) \binom{n}{t+2}p^{t+2}
\end{align}
while for asymmetric codes we have  
\begin{align}
\label{eq:RhoLAsymApprox}
    p_\mathrm{L} 
&\approx  (1-\beta_{e_\mathrm{g}+e_\mathrm{Z}+1}) \binom{n}{e_\mathrm{g}+e_\mathrm{Z}+1} p^{e_\mathrm{g}+e_\mathrm{Z}+1} 
\notag \\
&+ (1-\beta_{e_\mathrm{g}+1}) \binom{n}{e_\mathrm{g}+1} p^{e_\mathrm{g}+1} 
\end{align}
where $e_\mathrm{g} = \min\left\{t_\mathrm{X}, t_\mathrm{Z}\right\}$ and $e_\mathrm{Z} = \max \left\{0,t_\mathrm{Z} - e_\mathrm{g}\right\}$ for a $[[n, k,d_\mathrm{X}/d_\mathrm{Z}]]$ code \cite{ChiVal:20a}.
Thus, for a symmetric quantum error correcting code with $t=\lfloor(d-1)/2\rfloor$, once we have the values of $f_{t+1}(i,\ell)$, by using \eqref{eq:betaA} we can write the analytical performance expression \eqref{eq:RhoLSymApprox} which is valid over an asymmetric channel with arbitrary asymmetry $A$, including the depolarizing channel ($A=1$). 
Similarly, for asymmetric codes with correction capability $e_\mathrm{g}$ and $e_\mathrm{Z}$, once we have $f_{e_\mathrm{g}+1}(i,\ell)$ and $f_{e_\mathrm{g}+e_\mathrm{Z}+1}(i,\ell)$ we can evaluate the performance over arbitrary asymmetric channels by using \eqref{eq:betaA} and \eqref{eq:RhoLAsymApprox}.

The values of $f_{j}(i,\ell)$ of interest can be obtained,  for the surface codes  analyzed in this paper, with limited complexity by a decoder error patterns search, as illustrated in Section~\ref{sec:NumRes}. Equivalently, we can derive the values of $\beta_{j}$ of interest starting from the knowledge of the logical operators weight distribution. %, aided by a preliminary error weight enumeration obtained via MacWilliam identities.
%We use the MacWilliam identities to provide some insights on the codes under examination, while the values of $f_{j}(i,\ell)$ will be obtained by a decoder error patterns search.
To this aim, we define the undetectable errors weight enumerator function, for a $[[n,k,d]]$ quantum code, as 
\begin{align}
\label{eq:deflogicals_eq}
L(z) &= \sum_{w = 0}^{n} L_w z^w
\end{align} 
where $L_w$ is the number of undetectable errors (logical operators) of weight $w$ \cite{ForValChi:23}. 
This polynomial can be easily computed starting from the code's generators by using the MacWilliams identities \cite{ForValChi:23}. From it, we can evaluate the errors compatible with the exact number of logical operators of a certain weight $w$, leading to the $\beta_{j}$.

\section{Quantum Topological Planar Codes} \label{sec:quantum_top}

\begin{figure*}[t]
	\centering
	\includegraphics[width=\textwidth]{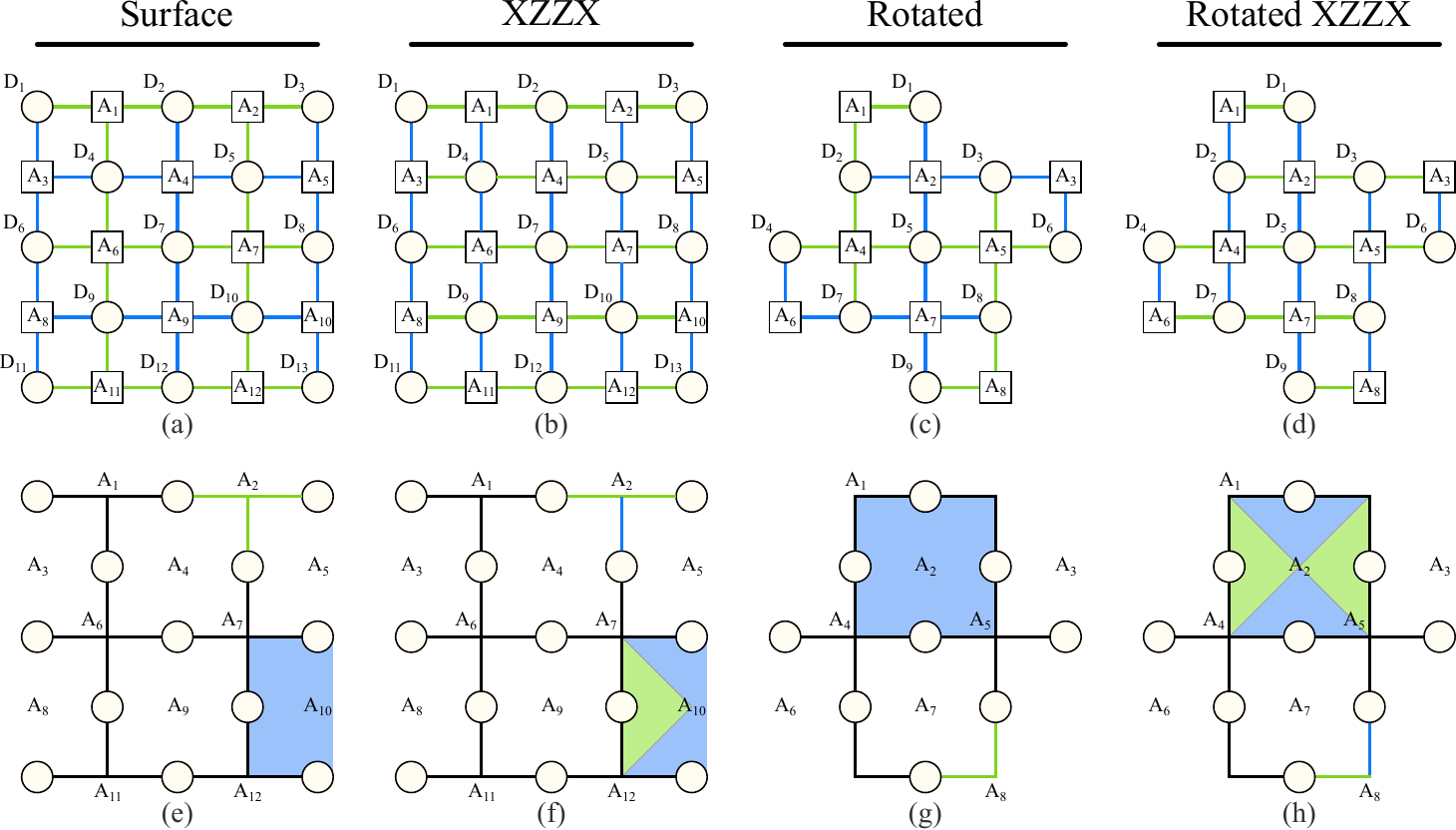}
	\caption{(a, e) $[[13,1,3]]$ surface code, (b, f) $[[13,1,3]]$ XZZX code, (c, g) $[[9,1,3]]$ rotated surface code, (d, h) $[[9,1,3]]$ rotated XZZX code. The first row represent the actual code structure with data qubits (circles) and ancillas (squares). $\M{Z}$ measurements are depicted in blue, while $\M{X}$ measurements are depicted in green. The second row is a simplified representation of the lattice with sites and plaquettes.}
	\label{Fig:Codici}
\end{figure*}

In this section, we analyze some topological codes which can be mapped into planar structures and measured using local operations (i.e., error correction is performed letting interact qubits which are physically nearby). 
These codes are of practical interest for applications where system scalability is required.
In the following, we provide analytical tools that can be used to reproduce the performance of such codes in any polarizing quantum channel (i.e., given $p_\mathrm{X}$, $p_\mathrm{Y}$, and $p_\mathrm{Z}$).

In Fig.~\ref{Fig:Codici}a-d  we show the actual physical structure of some planar codes.
For example, in Fig.~\ref{Fig:Codici}a we depict the $[[13,1,3]]$ surface code. 
Here, a codeword is constituted by $n=13$ qubits, represented with circles. 
The error syndrome is extracted by measuring the stabilizers of the code \cite{Got:09}. In particular, the measurement is performed by letting interact codeword qubits with auxiliary qubits, usually referred as ancillas, which are indicated with squares in the figure \cite{NieChu:10}. After the interaction, by measuring the ancilla qubits we obtain a syndrome composed by $n-k = 12$ bits.
Maximum likelihood decoders can be implemented using lookup tables. However, due to the structure of these codes, suboptimal decoder based on the \ac{MWPM} algorithm have been proposed \cite{Hig:22}. This decoder builds a graph where vertices correspond to error ancillas, and edges are weighted according to the number of qubits between them. Then, it finds the shortest path between each couple of ancillas.  
For example, considering the code represented in Fig.~\ref{Fig:Codici}a, if the error $\M{Z}_7$ occurs, $A_6$ and $A_7$ detect an error in their neighbourhood. Finally, connecting these two ancillas the \ac{MWPM} is able to localize the error.
Due to degeneracy of these quantum codes, the decoder based on \ac{MWPM} is able to guarantee the error correction capability $t$ and is also able to correct some error patterns of weight greater than $t$.

Before proceeding with the analysis of topological codes, we introduce another graphical notation and some terminologies to describe them. 
In fact, these codes are usually represented in terms of a planar lattice where vertices and cells are called \emph{sites} and \emph{plaquettes}, respectively.
In Fig.~\ref{Fig:Codici}e-h we sketch the graphical representation which highlights sites and plaquettes. 
For the sake of clarity, in Fig.~\ref{Fig:Codici}e a site named $A_2$ is highlighted in green, while a plaquette named $A_{10}$ is highlighted in blue.
Sites and plaquettes represent ancilla qubits in this notation and therefore perform Pauli measurements on the adjacent \emph{edges} which stand for physical qubits and are represented with circles in the figure. 
Moreover, edges (i.e., qubits) which resides on the boundary of the lattice can be categorized as \emph{smooth} and/or \emph{rough}.
They are defined as smooth, when the perpendicular ancilla measurement to the boundary is a Pauli $\M{X}$ measurement, and are defined as rough, when the perpendicular ancilla measurement to the boundary is a Pauli $\M{Z}$ measurement.
For example, in Fig.~\ref{Fig:Codici}e the boundary qubit $D_2$ is rough and $D_3$ is both rough and smooth (above is rough and on the right is smooth).
This notation is used during the decoding procedure when ghost ancillas (i.e., virtual ancillas not physically implemented) are employed.

\begin{figure*}[t]
	\centering
 \includegraphics[width=\textwidth]
 {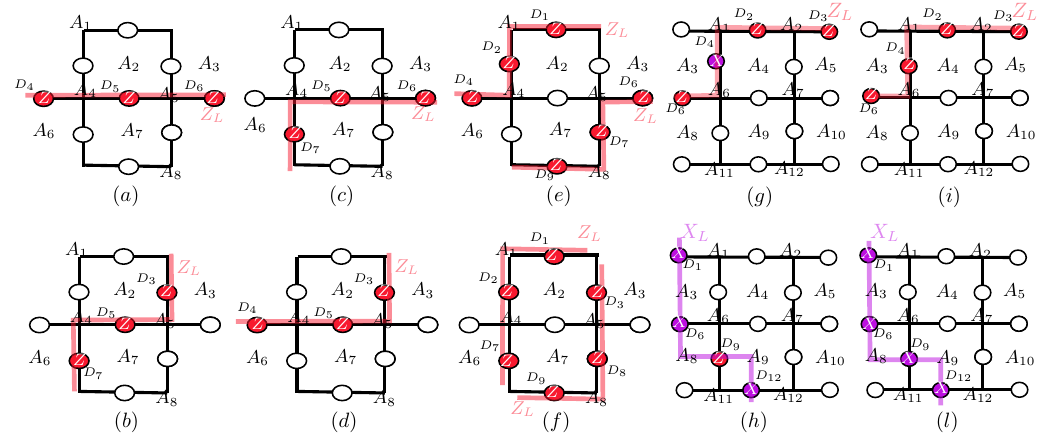}
 
	%\resizebox{0.99\textwidth}{!}{%\input{Figures/surface_rot_LOG.txt}
	%}
	\caption{ (a-f) Examples of $\M{Z}$ logical operators of weight $w = 3$ of the $[[ 9,1,3 ]]$ rotated surface code. Data qubits with a $\M{Z}_L$ error are depicted in red. The last four $\M{Z}$ logical operators have the form of (e) and (f) but with $\M{Y}$ operators on qubits $D_1$, $D_2$, $D_8$ and $D_9$. The other 12 $\M{X}$ logical have the same structure on the dual lattice. (g,h) Examples of $\M{Z}_L$ and $\M{X}_L$ logical operators of weight $w = 4$ for the $[[ 13,1,3 ]]$ XZZX code. (i,l) Examples of $\M{Z}_L$ and $\M{X}_L$ logical operators of weight $w = 4$ for the $[[ 13,1,3 ]]$ surface code. }
	\label{Fig:sur_rot_log}
\end{figure*}

\subsection{Surface Codes} 
\label{subsec:surface} In this subsection we describe the original surface codes and revise some recent results on their performance \cite{ForValChi:23}, which will serve for the comparison with the surface codes variants.
Surface codes are topological codes that can be mapped into a planar rectangular $d_\mathrm{X} \times d_\mathrm{Z}$ lattice. For example, in Fig.~\ref{Fig:Codici}e is shown a lattice with $d_\mathrm{X} = d_\mathrm{Z} = 3$. 
%Such codes were the first quantum planar codes to be proposed in literature \cite{BraKit:98,DenKitLan:02}. 
A symmetric surface code with distance $d = d_\mathrm{X} = d_\mathrm{Z}$ has $d^2 + (d-1)^2$ edges, which correspond to physical qubits \cite{Rof:19}. 
This structure has $d \cdot (d-1)$ sites and $d \cdot (d-1)$ plaquettes, which are associated with $\M{X}$ generators, and $\M{Z}$ generators, respectively.
This means that each ancilla measurement uses only operators of one type, $\M{Z}$ or $\M{X}$.
%In particular, generators in the interior are four-qubits operators, while, along plaquette (or rough) and site (or  smooth) edges, generators are $\M{Z}^{\otimes 3}$ and  $\M{X}^{\otimes 3}$ operators, respectively.
Moreover, due to the physical structure of the code, generators in the inner part of the lattice are four-qubits operators, while along the boundaries we have three-qubits operators.
For this reason, they can be interpreted also as \ac{QLDPC} codes~\cite{MacMitMcF:04,OstOrsLaz:22}. 

As regards logical operators, $\M{Z}_L$ can be seen as a tensor product of $\M{Z}$'s acting on a chain of qubits running from a rough edge to one at the opposite side of the lattice. 
In a similar way, the $\M{X}_L$ will be the tensor product of $\M{X}$'s on a chain running from a smooth edge to one at the other side of the dual lattice. In the case described so far, the lattice has a square shape, as shown in Fig.~\ref{Fig:Codici}a/e which represents the smallest surface code, with parameters $[[13,1,3]]$. 
Symmetric surface codes with higher distances can be easily built by adding the same number of columns and rows to this lattice. In this way, it is also possible to create asymmetric surface codes by adopting a rectangular lattice. For instance the $[[23,1,3/5]]$ can be obtained by adding two additional columns to the $[[13,1,3]]$. By doing so, the shortest $\M{Z}_L$ logical operator will be made of five $\M{Z}$ Pauli operators, while the shortest $\M{X}_L$ will still consist of three $\M{X}$ Pauli operators. Indeed this code has distances $d_\mathrm{X} = 3$ and $d_\mathrm{Z} = 5$. In general, asymmetric surface codes are characterized by parameters $[[d_\mathrm{X}d_\mathrm{Z} + (d_\mathrm{X}-1)(d_\mathrm{Z}-1), 1, d_\mathrm{X}/d_\mathrm{Z}]]$.

For the $[[13,1,3]]$ surface code the $L(z)$ has been computed in \cite{ForValChi:23},  obtaining $L(z) = 6 z^3 + 24 z^4 + 75 z^5 + 240 z^6 + 648 z^7 + 1440 z^8 + 2538z^9 + 3216z^{10} + 2634z^{11} + 1224z^{12} + 243z^{13}$. 
From this, we see that there are $6$ logical operators of weight $w = 3$. Thus, for the evaluation of $\beta_2$ we can restrict our analysis to error patterns of weight $2$ combined with a single qubit error correction operator.
Since the $[[13,1,3]]$ surface code logical operators with $w=3$ are all composed by three $\M{X}$ or three $\M{Z}$ by construction, we have that channel errors triggering a failure have to be composed by two $\M{X}$ or two $\M{Z}$. 
Note that an error pattern of the type $\M{X}\M{Y}$ has two $\M{X}$ operators, due to the fact that $\M{Y}$ can be decomposed as an $\M{X}$ and a $\M{Z}$ operator.
Then, for each logical operator, there are $\binom{3}{2}=3$ possible ways to have $2$ errors distributed in $3$ possible locations (i.e., the locations given by the logical operator).
Furthermore, when the logical error under examination is $\M{X}_L$ ($\M{Z}_L$), the error patterns that could generate it are in the form of $\M{X}\M{X}$ ($\M{Z}\M{Z}$), $\M{X}\M{Y}$ ($\M{Z}\M{Y}$), and $\M{Y}\M{Y}$. 
This results in $\binom{2}{2} + \binom{2}{1} + \binom{2}{2} = 4$ possible error configurations.
Then, we have in total $6\cdot3\cdot4$ error patterns of weight $2$ which produce logical operators of weight $3$ (i.e., a decoding failure).

The total number of logical operators with $w=4$ are $L_4 = 24$, but $8$ of them are considered implicitly. In particular, they are taken into account when considering that error patterns of the type $\M{Y}\M{Y}$ cause logical operators of weight $w=3$ \cite{ForValChi:23}. 
The remaining operators are $8$ $\M{Z}_L$ composed only by $\M{Z}$, and $8$ $\M{X}_L$ composed only by $\M{X}$. 
Logical operators with $w=4$ are more difficult to treat due to the fact that they could share $j$ or more Pauli operators. 
To simplify the discussion we consider pairs of logical operators. 
Let us consider for example the logical operator depicted in Fig.~\ref{Fig:sur_rot_log}i, $\M{Z}_L=\M{Z}_2\M{Z}_3\M{Z}_4\M{Z}_6$ and the operator $\M{Z}_L=\M{Z}_3\M{Z}_5\M{Z}_6\M{Z}_7$.
For a given operator with $w=4$, we have $\binom{4}{2}=6$ possible patterns of $j=2$ Pauli errors.
However, we can see that two of these six configurations never cause the logical operator $\M{Z}_L=\M{Z}_2\M{Z}_3\M{Z}_4\M{Z}_6$. In particular, $\M{Z}_4\M{Z}_6$ would be corrected due to the degeneracy of the code, while $\M{Z}_2\M{Z}_3$ would cause a different logical operator $\M{Z}_L=\M{Z}_1\M{Z}_2\M{Z}_3$ (a configuration that we have already counted using $w=3$). 
Hence, we have to subtract these two patterns from the total of $\binom{4}{2}=6$.
In addition, we have to consider that $\M{Z}_2\M{Z}_4$ produce the same error syndrome of $\M{Z}_3\M{Z}_6$ (the same happens for $\M{Z}_2\M{Z}_6$ and $\M{Z}_3\M{Z}_4$).
Since the decoder acts in a deterministic way, only one of these two must be counted.
However, $\M{Z}_3\M{Z}_6$ could trigger an error also for $\M{Z}_L=\M{Z}_3\M{Z}_5\M{Z}_6\M{Z}_7$.
In the end, we have that each of the $8$ pairs of logical operators with $w=4$ can be caused only by $3$ error patterns.
We conclude that $\beta_2$ for the $[[13,1,3]]$ code with \ac{MWPM} decoding over a depolarizing channel ($A=1$) is
%From this polynomial we can analytically compute $\beta_2$ for \ac{MWPM} decoding over a depolarizing channel ($A=1$), giving
%
%\begin{align} 
%\label{eq:Beta2Surf}
 %   \beta_2 &= 1 - 3 \frac{2\, L_3(2,0) + 4\, L_3(1,0) + L_3(0,0) }{\binom{13}{2} 3^2} \notag \\&- 1.5 \frac{2\, L_4(2,0) + 4\, L_4(1,0) + L_4(0,0) }{\binom{13}{2} 3^2} \notag \\ &= 1 - \frac{6\cdot 3 + 12\cdot 3 + 3\cdot 6 + 3\cdot 8 + 6\cdot 8 + 1.5\cdot 16 }{\binom{13}{2} 3^2} \notag\\ &= \frac{89}{117} \simeq 0.76.
%\end{align}

% \begin{align}
% \label{eq:Beta2Surf}
%     \beta_2 = 1 - \frac{6\cdot3\cdot4 + 16\cdot\frac{3}{2}\cdot4}{\binom{13}{2} 3^2} = \frac{89}{117} \simeq 0.76.
% \end{align}
% %
\begin{align}
\label{eq:Beta2Surf}
    \beta_2 = 1 - \frac{6\cdot3\cdot4 + 8\cdot3\cdot4}{\binom{13}{2} 3^2} = \frac{89}{117} \simeq 0.76.
\end{align}
This result was derived in \cite{ForValChi:23}, but notably there are no analyses in the literature for the rotated variant as well as the XZZX surface codes. Hence, in the next subsections we start from the undetectable error weight enumerator $L(z)$ and we provide new expressions for the  rotated and XZZX codes.

%For the sake of clarity, we show now the reasoning behind the evaluation of the term $6\cdot3\cdot4$. Firstly, from $L(z)$ we see that there are six logical operators of weight $w = 3$. Each one of them can be caused by three different patterns of Pauli error of weight $j = 3$, and we must take into account errors of the kind $\M{Z}$$\M{Z}$ ($\M{X}$$\M{X}$), $\M{Z}$$\M{Y}$ ($\M{X}$$\M{Y}$), $\M{Y}$$\M{Z}$ ($\M{Y}$$\M{X}$), and $\M{Y}$$\M{Y}$, when considering $\M{Z}_L$ and $\M{X}_L$ operators, respectively. Note also that eight of the 24 logical operators of weight $w = 4$ are taken into account implicitly, including all patterns of $\M{Y}$$\M{Y}$ errors among the Pauli operators that generate logical operators of weight $w =3$. A further detailed explanation can be found in \cite{ForValChi:23}. 

\subsection{Rotated Surface Codes} 

%These codes are obtained starting from standard surface codes by removing $(d_\mathrm{X}-1)\cdot(d_\mathrm{X}-1)$ number of qubits and ancillas \cite{HorFowDev:12}. 
%This operation leads to a $[[d_\mathrm{X}d_\mathrm{Z}, 1, d_\mathrm{X}/d_\mathrm{Z}]]$ quantum code.
These codes are obtained starting from standard surface codes by removing $(d-1)^2$  codeword qubits \cite{HorFowDev:12}. 
Then, the structure is usually rotated by $45^\circ$ for graphical representation.
For this reason, literature refers to these as ``rotated'' surface codes, even if from a coding theory perspective we could call them ``punctured'' surface codes.
This procedure can be applied also to asymmetric codes, by removing $(d_\mathrm{X}-1)\cdot(d_\mathrm{Z}-1)$ codeword qubits from the asymmetric surface code, leading to a $[[d_\mathrm{X}d_\mathrm{Z}, 1, d_\mathrm{X}/d_\mathrm{Z}]]$ quantum code.

Let us take as an example the $[[13,1,3]]$ original surface code. After the puncturing, it becomes the $[[9,1,3]]$ rotated surface code depicted in Fig.~\ref{Fig:Codici}c/g. This is the code that has been implemented in \cite{KriSebLac:22,ZhaYouYe:22}. By following the MacWilliams identities approach \cite{ForValChi:23} we derive the \ac{WE} function for the $[[9,1,3]]$ rotated surface code as
\begin{equation}\label{eq:Lz913}
L(z) = 24 z^3 + 192 z^5 + 408 z^7 + 144 z^9 \,.
\end{equation}
In this case, we have $24$ logical operators of weight $w = 3$ and zero of weight $w = 4$. However, as we did for the original surface code when $w=4$, we consider only 16 out of 24 logical operators, due to the fact that we are implicitly counting the $8$ operators having both Pauli $\M{Z}$ and $\M{X}$ into the $16$ ones.
%the ones with only $\M{Z}$ or $\M{X}$ Pauli operators. 
%The last eight operators (constituted of both Pauli $\M{Z}$ and $\M{X}$) are taken into account implicitly. 
In fact, focusing for example on $\M{Y}_1\M{Y}_2\M{Z}_4$ for the code in Fig.~\ref{Fig:sur_rot_log}e, we observe that $\M{X}_1\M{X}_2$ is corrected by the decoder, resulting in the logical operator $\M{Z}_L=\M{Z}_1\M{Z}_2\M{Z}_4$.  
Let us focus on the 12 logical operators $\M{Z}_L$ of weight $w = 3$, and, in particular, on the eight constituted of only Pauli $\M{Z}$ operators, reported pictorially in Fig.~\ref{Fig:sur_rot_log}a-f. 
As shown in Section~\ref{subsec:surface}, we are interested in the number of $\M{Z}$ errors of weight $j = 2$ that can cause these logical operators. 
However, we can see that different logical operators share common patterns of $\M{Z}$ errors with weight $j = 2$ and must be considered only once. 
For instance, the operator in Fig.~\ref{Fig:sur_rot_log}a can be caused by three different patterns of $\M{Z}$ errors: $\M{Z}_4\M{Z}_5$, $\M{Z}_4\M{Z}_6$ and $\M{Z}_5\M{Z}_6$. In particular, $\M{Z}_4\M{Z}_5$ can cause either $(a)$ or $(d)$, while $\M{Z}_5\M{Z}_6$ can lead either to $(a)$ or to $(c)$, depending on the \ac{MWPM} implementation. Specifically, each logical operator of the kind (a-d) can be caused by two patterns of Pauli errors of weight $j=2$ which are in common with others. On average, each of these four operators can be caused by $\binom{3}{2} - 1 = 2$ different patterns of errors of weight $j=2$. As regards the four logical operators in Fig.~\ref{Fig:sur_rot_log}e and Fig.~\ref{Fig:sur_rot_log}g, they have just one pattern of errors of weight $j=2$ in common with each other. Hence, considering them in pairs as done in Section~\ref{subsec:surface}, we must take into account $ 2 \cdot  \binom{3}{2} - 1 = 5$ different pattern of errors per couple. %, which means $2.5$ patterns for the single logical operator. 
%As a consequence, $ \mu^{3}_2=\frac{(2\cdot4 + (5\cdot2)}{8} = 2.25$, and the value of $\alpha^{(3)}_2 = \frac{2.25}{\binom{3}{2}\binom{2}{2}\binom{0}{2}} = 0.75$}. 
Note that, these eight $\M{Z}_L$ logical operators can be caused by $\M{Z}\M{Z}$, $\M{Z}\M{Y}$, and $\M{Y}\M{Y}$ errors, since the additional $\M{X}_L$ errors are corrected by the code. Therefore, for the $[[9,1,3]]$ rotated surface code with \ac{MWPM} decoding over the depolarizing channel we have 
%
%\begin{align}
%\label{eq:Beta2Surf_rot}
  %  \beta_2 &= 1 - 2.25 \frac{2\, L_3(2,0) + 4\, L_3(1,0) + L_3(0,0)}{\binom{9}{2} 3^2}% - \mu \frac{L_3(0,0)}{\binom{9}{2} 3^2} 
    % \beta_2 &= 1 - 2 \frac{L_3(2,0)\cdot\mu_2(2,0)+ 
    % L_3(1,0)\cdot\mu_2(1,0)}{\binom{9}{2} 3^2} \notag\\
    % &-  \frac{L_3(0,0)\cdot\mu_2(0,0)}{\binom{9}{2} 3^2} 
    %&= 1 - \frac{8 \cdot 2.25 + 8 \cdot 2.25 + 16 \cdot 4.5 + 8 \cdot 2.25}{\binom{9}{2} 3^2} 
  % \notag \\ &= 1 - \frac{4.5\cdot 8 + 9 \cdot 8 + 2.25 \cdot 16}{\binom{9}{2} 3^2} = \frac{180}{324} \simeq 0.56.
%\end{align}
%
% %
% \begin{align}
% \label{eq:Beta2Surf_rot}
%     \beta_2 = 1 - \frac{(8 \cdot 2 + 8 \cdot 2.5) \cdot 4}{\binom{9}{2} 3^2} = \frac{180}{324} \simeq 0.56.
% \end{align}
% %
%
\begin{align}
\label{eq:Beta2Surf_rot}
    \beta_2 = 1 - \frac{(8 \cdot 2 + 4 \cdot 5) \cdot 4}{\binom{9}{2} 3^2} = \frac{5}{9}%\frac{180}{324} 
    \simeq 0.56.
\end{align}
It is also possible to derive $\beta_2$ over a phase flip channel, observing that in this case we have only $\M{Z}$ Pauli errors. This leads for the same $[[9,1,3]]$ rotated surface code over a phase flip channel to
%\begin{align}
%\label{eq:Beta2Surf_rot_ph}
   % \beta_2 = 1 - \frac{2.25 \cdot L_3(2,0)}{\binom{9}{2}} = \frac{18}{36} \simeq 0.5.
%\end{align}
%
% %
% \begin{align}
% \label{eq:Beta2Surf_rot_ph}
%     \beta_2 = 1 - \frac{4 \cdot 2 + 4 \cdot 2.5}{\binom{9}{2}} = \frac{18}{36} \simeq 0.5.
% \end{align}
% %
%
\begin{align}
\label{eq:Beta2Surf_rot_ph}
    \beta_2 = 1 - \frac{4 \cdot 2 + 2 \cdot 5}{\binom{9}{2}} = \frac{1}{2} %\frac{18}{36} 
    = 0.5.
\end{align}
From this analysis we observe that the rotated surface $[[9,1,3]]$ code has a worst $\beta_{t+1}$ parameter compared to the original $[[13,1,3]]$ code. 
However, it has a smaller number of qubits that can be affected by noise and, therefore, looking at $\beta_{t+1}$ is not sufficient to say which is optimal in terms of logical error rates.
In general, to find which is the best among two codes $\mathcal{C}_1$ and $\mathcal{C}_2$ we can compute the ratio between their asymptotic approximations 
\begin{align}
\label{eq:RatioAdvantage}
    r(p, A) = \frac{\binom{n_1}{t_1+1}\left[1-\beta_{t_1+1}(A)\right]}{\binom{n_2}{t_2+1}\left[1-\beta_{t_2+1}(A)\right]} p^{t_1-t_2}\,.
\end{align}
From $f_j(i, \ell)$ we can evaluate the $\beta_{t+1}(A)$ of the codes through \eqref{eq:betaA}, which allows us to analytically study the function $r$ against $A$ for comparative performance analysis. 
Considering as $\mathcal{C}_1$ the $[[13,1,3]]$ code and as $\mathcal{C}_2$ the $[[9,1,3]]$ one, we obtain from \eqref{eq:Beta2Surf} and \eqref{eq:Beta2Surf_rot} for the depolarizing channel that $r=1.18 > 1$, meaning that $\mathcal{C}_2$ is better than $\mathcal{C}_1$ over that channel.
Note that, in this case, $r$ is independent of $p$, indicating that the rotated $[[9,1,3]]$ code is uniformly better than the original $[[13,1,3]]$ surface code.

\subsection{XZZX Surface Codes} 

XZZX codes have the same structure as surface codes, i.e., the same amount of codeword qubits, ancillas, and generator weights, but with a different composition of the generators \cite{AtaTucBar:21}.
While in the original surface codes, each generator is composed only by $\M{X}$ or $\M{Z}$, in XZZX codes the generators inside the lattice are of the kind $\M{X}\M{Z}\M{Z}\M{X}$, and along smooth and rough boundaries we have $\M{X}\M{X}\M{Z}$ and $\M{Z}\M{Z}\M{X}$ generators, respectively. 
This can be also interpreted as a Hadamard rotation on some qubits of the lattice. 
For example, if we apply the Hadamard rotation to $D_4$, $D_5$, $D_9$, and $D_{10}$ in Fig.~\ref{Fig:Codici}a, we obtain the XZZX code of Fig.~\ref{Fig:Codici}b.
Since in surface codes half of the generators are composed of $\M{X}$ operators while the other half of $\M{Z}$ operators, errors on adjacent qubits produce horizontal or vertical error chains for both $\M{X}$ and $\M{Z}$ Pauli operators. 
On the contrary, XZZX codes have an intrinsic symmetry: $\M{X}$ errors can align only in vertical directions, while there are only horizontal $\M{Z}$ error chains. 
Due to this structure, we show that these codes work better than surface codes over asymmetric channels.%, where one kind of Pauli error happens more frequently than the others. 

We can compute for the $[[13,1,3]]$ XZZX code 
\begin{align}
L(z)  = \, & 6 z^3 + 24 z^4 + 75 z^5 + 240 z^6 + 648 z^7  \notag \\ 
 &  + 1440 z^8 + 2538z^9 + 3216z^{10} \notag \\ 
 & + 2634z^{11} + 1224z^{12} + 243z^{13} \, . 
\end{align}
As anticipated, this is the same as that obtained for the original surface code. 
Indeed, the logical operators have the same structure but they are composed of different Pauli operators. In particular, notice that, even if the logical operators differ for some Pauli operators with respect to the surface codes, a decoder will make its decision based only on the resulting syndrome. As a result, over the depolarizing channel and for a \ac{MWPM} decoder, for the XZZX variant of the $[[13,1,3]]$ code, we still have $\beta_2\simeq 0.76$, as can be checked by deriving it in the same way as \eqref{eq:Beta2Surf}.   
%
 %\begin{align}
 %\label{eq:Beta2XZZX}
  %   \beta_2 = 1 - \frac{6\cdot3\cdot4 + 16\cdot\frac{3}{2}\cdot4}{\binom{13}{2} 3^2} = \frac{89}{117} \simeq 0.76.
% \end{align}
%

Moving to the analysis over asymmetric channels, we remark that in the $[[13,1,3]]$ surface code the 16 logical operators which produce unique errors are composed by only $\M{Z}$ or only $\M{X}$ operators \cite{ForValChi:23}, while, for the $[[13,1,3]]$ XZZX code they consist of both Pauli operators. 
As a consequence, some of these operators occur only in the presence of both $\M{Z}$ and $\M{X}$ errors.
Considering an asymmetric channel where $\M{X}$ errors are less probable than $\M{Z}$ errors, it follows that logical operators of the kind shown in Fig.~\ref{Fig:sur_rot_log}g are, on average, less probable to occur.
This is the reason behind the XZZX performance advantage over asymmetric channels. 
For example, letting $A \to \infty$ (i.e., assuming phase flip channel), we have four error patterns of the type $\M{Z}$$\M{X}$ that we have to exclude, since $\M{X}$ errors cannot occur on this channel. Hence, over the phase-flip channel, the XZZX variant leads to 
% %
% \begin{align}
% \label{eq:Beta2XZZXph}
%     \beta_2 = 1 - \frac{3\cdot3 + 8\cdot\frac{3}{2} - 4}{\binom{13}{2} } = \frac{61}{78} \simeq 0.782.
% \end{align}
% %
%
\begin{align}
\label{eq:Beta2XZZXph}
    \beta_2 = 1 - \frac{3\cdot3 + 4\cdot3 - 4}{\binom{13}{2} } = \frac{61}{78} \simeq 0.782.
\end{align}
This shows how the error correction capability of XZZX codes increases with the asymmetry of the channel. More results will be illustrated for other values of $A$ in Section~\ref{sec:NumRes}.  

\subsection{Rotated XZZX Codes} 

\begin{figure}[t]
	\centering
	\includegraphics[width=\columnwidth]{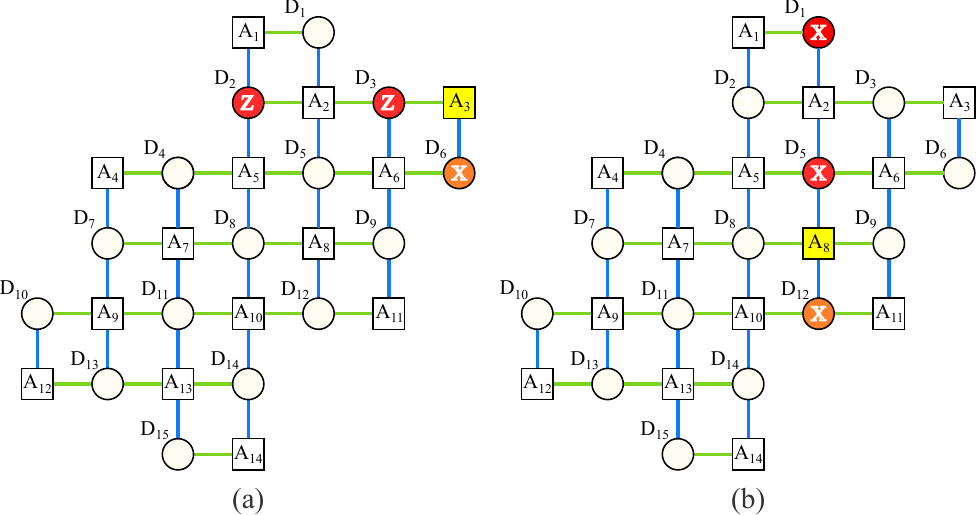}
	\caption{Examples of undetected error patterns on the rotated XZZX $[[15,1,3]]$. In red are represented errors introduced by the channel, in yellow the ancilla that measures ``-1'' during the syndrome extraction, and in orange the error correction applied by the \ac{MWPM} decoder when the current highlighted ancilla is given as input. (a) A particular $\M{Z}\M{Z}$ error producing a logical error when decoded (unresolvable error). (b) A particular $\M{X}\M{X}$ error producing a logical error when decoded. %For this reason this code does not guarantee 
 }
	\label{Fig:RotXZZX_UnresolvableErrors}
\end{figure}

Similarly to the XZZX, these codes can be constructed starting from the rotated surface code (see Fig.~\ref{Fig:Codici}c) and then letting the ancillas measure $\M{Z}$ in the horizontal direction, and $\M{X}$ in the vertical direction (see Fig.~\ref{Fig:Codici}d).
Thus, for a symmetric structure, we obtain a $[[d^2, 1, d]]$ error correcting code. 
For symmetric structures, this procedure allows reducing the number of physical qubits without changing the distance of the code. 
For example, the $[[9,1,3]]$ and the $[[25,1,5]]$ rotated XZZX codes are those implemented as  described in \cite{AchRajAle:22}.  
In this way, for a bounded distance decoder that corrects error patterns with up to  $t$ errors, we have that rotated codes are better in terms of performance, due to the reduction of the codeword length. 
However, for complete decoders, this is not always true due to the fact that, as shown in \eqref{eq:RatioAdvantage}, the performance gap depends on both the codeword length and the values of $\beta_j$.
We can compute for the $[[9,1,3]]$ rotated XZZX code the \ac{WE} function obtaining
\begin{align}
L(z) = 24 z^3 + 192 z^5 + 408 z^7 + 144 z^9
\end{align}
which, as expected, is the same as the non-XZZX version \eqref{eq:Lz913}. For the depolariziong channel, the resulting $\beta_2=5/9$ is, therefore, equal to \eqref{eq:Beta2Surf_rot} computed for the $[[9,1,3]]$ rotated surface code. Thus, the XZZX variant does not change the performance over the depolarizing channel. However, the codes perform differently on asymmetric channels. In fact, let us now consider the 18 configurations of errors with weight $w = 2$ of \eqref{eq:Beta2Surf_rot_ph}, which, in the rotated surface code, correspond to only $\M{Z}$ Pauli errors. As we did in Fig.~\ref{Fig:sur_rot_log} for non-rotated codes, it can be shown that 10 of these configurations include one $\M{X}$ Pauli error in the case of the $[[9,1,3]]$ rotated XZZX code. As a consequence, this code has a great advantage over asymmetric channels. For example, we compute for it   
% %
% \begin{align}
% \label{eq:Beta2Surf_rot_ph}
%     \beta_2 = 1 - \frac{4 \cdot 2 + 4 \cdot 2.5 - 10}{\binom{9}{2} } = \frac{28}{36} \simeq 0.778
% \end{align}
% %
%
\begin{align}
\label{eq:Beta2Surf_rot_ph}
    \beta_2 = 1 - \frac{4 \cdot 2 + 2 \cdot 5 - 10}{\binom{9}{2} } = \frac{7}{9} %\frac{28}{36} 
    \simeq 0.778
\end{align}
over a phase flip channel, to be compared with $\beta_2=1/2$ of the non-XZZX code given in \eqref{eq:Beta2Surf_rot}. 

Finally, we could be tempted to apply the XZZX variant to rectangular rotated surface codes. However, it is easy to see that this will equalize the $d_\mathrm{X}$ and $d_\mathrm{Z}$ distances, so losing correction capabilities.   
In fact, if we start from an asymmetric structure of dimension $d_\mathrm{X} \times d_\mathrm{Z}$, with parameters $[[d_\mathrm{X}d_\mathrm{Z}, 1, d_\mathrm{X}/d_\mathrm{Z}]]$ when we apply the XZZX variant the code becomes $[[d_\mathrm{X}d_\mathrm{Z}, 1, \min\{d_\mathrm{X},d_\mathrm{Z}\}]]$.
This is due to some error patterns which reduce the error correction capability of the starting code. 
We report in Fig.~\ref{Fig:RotXZZX_UnresolvableErrors}a an example where, starting from the $[[15,1,3/5]]$ rotated code, we apply the XZZX variant. Here we show a decoding error caused by an $\M{Z}\M{Z}$ error pattern. Since a weight two pattern is not corrected, the code distance is not $d_\mathrm{Z}=5$ for $\M{Z}$ errors.
In Fig.~\ref{Fig:RotXZZX_UnresolvableErrors}b we also show a failure caused by an $\M{X}\M{X}$ error, which confirms that this asymmetric lattice with the XZZX variant has a symmetric distance $d=3$.
In general, this is the reason why the surgery on the lattice which transforms a squared structure to a rectangular one is not able to improve the distance on rotated XZZX codes.

\section{Numerical Results}\label{sec:NumRes}

\subsubsection{Error patterns search via decoding} 

\begin{table*}[t]
    %\justifying
    \centering
    \setlength{\tabcolsep}{1pt}
    \caption{Fraction of non-correctable error patterns $f_j(i,\ell)$ per error class of several topological planar codes decoded by \acl{MWPM}: Surface, XZZX, Rotated surface and Rotated XZZX, having symmetric and asymmetric error correction capabilities.}
    \label{tab:Err}
    \small
    %\resizebox{0.99\textwidth}{!}{
    %\begin{tabular}{l  C{2.5cm} C{2.25cm} C{2.5cm} C{2.75cm}}
    \begin{tabular}{lC{1.3cm}C{1.3cm}C{1.3cm}C{1.3cm}C{1.3cm}C{1.3cm}C{1.3cm}C{1.3cm}C{1.3cm}C{1.3cm}}
        \toprule
        %\multirow{2}{*}{
        \rowcolor[gray]{.95}
       & & & & & \textbf{Surface} & & & & &\\
        \midrule
        \rowcolor[gray]{.95}
        \textbf{Code} & $\M{X}\M{X}$ & $\M{X}\M{Z}$ & $\M{X}\M{Y}$ & $\M{Z}\M{Z}$ & $\M{Z}\M{Y}$ & $\M{Y}\M{Y}$ & & & &\\
        \midrule
        $[[13,1,3]]$ & $0.27$ & $0$ & $0.27$  & $0.27$  & $0.27$ & $0.51$ \\
        $[[9,1,3]]$ & $0.5$ & $0$ & $0.5$  & $0.5$  & $0.5$ & $1$ \\
        $[[23,1,3/5]]$ & $0.174$ & $0$ & $0.174$ & $0$ & $0$ & $0.174$ \\
        $[[15,1,3/5]]$ & $0.324$ & $0$ & $0.324$ & $0$ & $0$ & $0.324$ \\
        \midrule 
        \rowcolor[gray]{.95}
        \textbf{Code} & $\M{X}\M{X}\M{X}$ & $\M{X}\M{X}\M{Z}$ & $\M{X}\M{X}\M{Y}$ & $\M{X}\M{Z}\M{Z}$ & $\M{X}\M{Z}\M{Y}$ & $\M{X}\M{Y}\M{Y}$ & $\M{Z}\M{Z}\M{Z}$ & $\M{Z}\M{Z}\M{Y}$ & $\M{Z}\M{Y}\M{Y}$ & $\M{Y}\M{Y}\M{Y}$\\
        \midrule
       % $[[13,1,3]]$ & $0.52$ & $0.27$ & $0.52$ & $0.27$ & $0.45$ & $0.67$ & $0.53$ & $0.53$ & $0.68$ & $0.78$ \\
        $[[23,1,3/5]]$ & $0.434$ & $0.174$ & $0.434$ & $0$ & $0.174$ & $0.434$ & $0.066$ & $0.066$ & $0.235$ & $0.487$ \\
        $[[15,1,3/5]]$ & $0.635$ & $0.324$ & $0.635$ & $0$ & $0.324$ & $0.635$ & $0.279$ & $0.279$ & $0.603$ & $0.868$ \\
        $[[41,1,5]]$ & $0.0212$ & $0$ & $0.0212$ & $0$ & $0$ & $0.0212$ & $0.0212$ & $0.0212$ & $0.0212$ & $0.042$ \\
        $[[25,1,5]]$ & $0.127$ & $0$ & $0.127$ & $0$ & $0$ & $0.127$ & $0.127$ & $0.127$ & $0.127$ & $0.254$ \\
\bottomrule
\toprule
           \rowcolor[gray]{.95}
       & & & & & \textbf{XZZX} & & & & &\\
        \midrule
        \rowcolor[gray]{.95}
        \textbf{Code} & $\M{X}\M{X}$ & $\M{X}\M{Z}$ & $\M{X}\M{Y}$ & $\M{Z}\M{Z}$ & $\M{Z}\M{Y}$ & $\M{Y}\M{Y}$ & & & &\\
        \midrule
        $[[13,1,3]]$ & $0.218$ & $0.051$ & $0.27$  & $0.218$  & $0.27$ & $0.513$ \\
        $[[9,1,3]]$ & $0.222$ & $0.278$ & $0.5$  & $0.222$  & $0.5$ & $1$ \\
        $[[23,1,3/5]]$ & $0.123$ & $0.016$ & $0.138$ & $0$ & $0.016$ & $0.154$ \\
        $[[15,1,3]]$ & $0.086$ & $0.086$ & $0.171$ & $0.067$ & $0.152$ & $0.324$ \\
        \midrule 
        \rowcolor[gray]{.95}
        \textbf{Code} & $\M{X}\M{X}\M{X}$ & $\M{X}\M{X}\M{Z}$ & $\M{X}\M{X}\M{Y}$ & $\M{X}\M{Z}\M{Z}$ & $\M{X}\M{Z}\M{Y}$ & $\M{X}\M{Y}\M{Y}$ & $\M{Z}\M{Z}\M{Z}$ & $\M{Z}\M{Z}\M{Y}$ & $\M{Z}\M{Y}\M{Y}$ & $\M{Y}\M{Y}\M{Y}$\\
        \midrule
       % $[[13,1,3]]$ & $0.52$ & $0.27$ & $0.52$ & $0.27$ & $0.45$ & $0.67$ & $0.53$ & $0.53$ & $0.68$ & $0.78$ \\
        $[[23,1,3/5]]$ & $0.281$ & $0.153$ & $0.311$ & $0.043$ & $0.18$ & $0.353$ & $0.043$ & $0.086$ & $0.248$ & $0.443$ \\
        $[[15,1,3]]$ & $0.244$ & $0.241$ & $0.399$ & $0.236$ & $0.391$ & $0.619$ & $0.211$ & $0.380$ & $0.391$ & $0.868$ \\
        $[[41,1,5]]$ & $0.014$ & $0.002$ & $0.016$ & $0.002$ & $0.007$ & $0.021$ & $0.013$ & $0.016$ & $0.021$ & $0.042$ \\
        $[[25,1,5]]$ & $0.027$ & $0.034$ & $0.06$ & $0.034$ & $0.067$ & $0.127$ & $0.027$ & $0.06$ & $0.127$ & $0.254$ \\
      
      %  \midrule 
        %\rowcolor[gray]{.95}
      %  & & \multicolumn{4}{c}{$1-\beta_2(A)$} & & \multicolumn{4}{c}{$1-\beta_3(A)$}  \\
      %  \cmidrule{3-6} \cmidrule{8-11}
      %  \rowcolor[gray]{.95}
     %   \textbf{Code}& & $A=1$ & $A=10$ & $A=100$ & $A\to \infty$ & & $A=1$ & $A=10$ & $A=100$ & $A\to \infty$  \\
        %\cmidrule{1-1} \cmidrule{3-6} \cmidrule{8-11}
      %  \midrule
     %   $[[13,1,3]]$ & & $0.24$ &$0.233$ &$0.265$ & $0.27$ & & $0.52$ &$0.48$ & $0.523$& $0.53$ \\
     %   $[[23,1,3/5]]$ & & $0.07$ & $0.0044$ & $6 \cdot 10^{-5}$& $0$ & & $0.203$ &$0.0736$ & 0.0778& $0.08$ \\
      %  $[[41,1,5]]$& & $0$&$0$ &$0$ &$0$ & & $0.014$ &$0.019$ &$0.023$ & $0.024$\\
      
        \bottomrule

    \end{tabular}
    %} $6.05\cdot 10^{-5}$
\end{table*}

In Tab.~\ref{tab:Err} we report the percentage of non-correctable errors $f_j(i,\ell)$ (i.e., composed by $j$ Pauli operator, $i$ of type $\M{Z}$ and $\ell$ of type $\M{X}$) for some surface, XZZX, rotated surface, and rotated XZZX codes.
These values have been evaluated by enumerating the error patterns of interest and running the \ac{MWPM} decoder. 
To this aim, we built a part of the decoder using the Lemon C++ library \cite{DezBalJut:11}, which provides an efficient implementation of graphs and networks algorithms. 
%Looking in the table, for the $[[13,1,3]]$ surface code, it results $f_2(0,2) = 0.27$. 
As anticipated, we can observe from Tab.~\ref{tab:Err} that each asymmetric code is able to correct all patterns of $\M{Z}\M{Z}$, apart from the rotated XZZX code using a rectangular lattice of dimension $3 \times 5$ which has a value of $f_2(2,0) \neq 0$. 
Note also that both the $[[9,1,3]]$ surface and XZZX codes are not able to correct any error patterns of the kind $\M{Y}\M{Y}$, being $f_2(0,0) = 1$. 
The table also highlights that, surface codes always correct error patterns composed by $i$ Pauli $\M{Z}$ and $\ell$ Pauli $\M{X}$, for all $i, \ell \le t$.
This is due to the fact that effective logical operators have only one kind of Pauli operator, which is no longer true in the case of XZZX codes. 

By examining Tab.~\ref{tab:Err}, it becomes apparent that the rotated XZZX surface code on a rectangular lattice exhibits a degradation in code distance. Specifically, without the application of the XZZX variant, the code is characterized as $[[15, 1, 3/5]]$, but upon applying the XZZX variant, it becomes a $[[15, 1, 3]]$ code. Notably, in this specific configuration, there is no enhancement of asymmetric error correction capability through the utilization of XZZX.

Finally, we stress the fact that by using these tabular values, we can write analytical expressions for the code performance.
This result can be used to analyze and design complex systems without implementing the decoder.
For example, the logical error rate of a rotated XZZX $[[9,1,3]]$ code tends for small $p$ to 
\begin{align}\label{eq:RhoL913XZZX}
    p_\mathrm{L} \approx \frac{0.222A^2 + 1.556A + 2.222}{(A + 2)^2} \, \binom{9}{2} p^2 \,.
\end{align}
Similarly, for the original $[[13,1,3]]$ surface code we have 
\begin{align}
    p_\mathrm{L} \approx \frac{0.27A^2 + 0.54A + 1.32}{(A + 2)^2} \, \binom{13}{2} p^2\,.
\end{align}
For example, for $A=10$ we obtain $p_\mathrm{L} \to 10 p^2$ for the rotated $[[9,1,3]]$ XZZX code and $p_\mathrm{L} \to 18.3 p^2$ for the $[[13,1,3]]$ surface code. 

\subsubsection{Asymptotic approximation and bounded distance decoder} 
\begin{figure}
	\centering
 \includegraphics[width=\columnwidth]{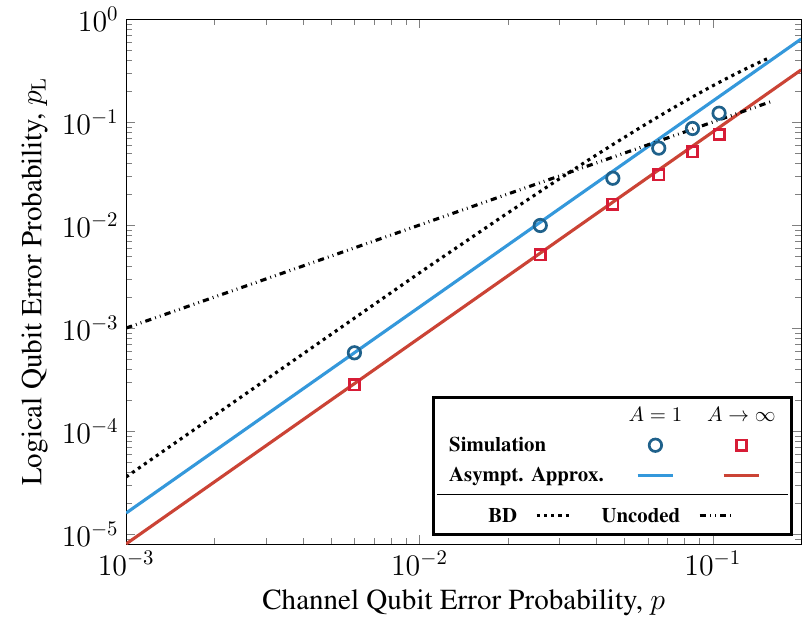}
	%\resizebox{0.99\columnwidth}{!}{
	    %\input{Figures/PerformanceSimulation}
	%}
	\caption{Logical error probability of the $[[9,1,3]]$ XZZX code over depolarizing ($A=1$) and phase flip ($A\to\infty$) channels. Comparison between simulation (mark symbols), asymptotic approximations (solid lines), and bounded distance decoding performance (dotted line).}
	\label{Fig:perf3d}
\end{figure}

In Fig.~\ref{Fig:perf3d} we report the asymptotic approximations computed using \eqref{eq:RhoL913XZZX} 
for the $[[9,1,3]]$ rotated XZZX code, over the depolarizing and phase flip channels. 
We notice that the estimates provided in \eqref{eq:RhoLSymApprox} and \eqref{eq:RhoLAsymApprox} closely align with the results obtained from simulations using \ac{MWPM} decoding.
This shows how the performance of an error-correcting quantum code with a complete decoder can be accurately described in the typical region of interest, i.e., $p < 0.1$, where the code is actually providing an improvement compared to the uncoded case.
Moreover, we show in the same figure the bounded distance decoding performance, obtained by  considering $\beta_j = 0$, for $j > t = \lfloor(d-1)/2\rfloor$.
The bounded distance decoding is unaffected by the asymmetry parameter. 
For this reason, it is not able to describe the advantage of XZZX codes over asymmetric channels. 
Specifically, the gap between bounded distance and \ac{MWPM} decoding is due to the fact that topological codes are able to correct a large number of errors of weight $w \geq t+1$.  
The $[[9,1,3]]$ XZZX code has the greatest error correction capability over the phase flip channel, where it can exploit its intrinsic symmetries which make one kind of Pauli error to align always in the same direction. 
%The plot shows also the bounded distance error probability for the $[[9,1,3]]$ XZZX code, which is not a function of the asymmetry of the channel. 
%

\subsubsection{Comparison of topological planar codes} 
\begin{figure}
	\centering
 \includegraphics[width=\columnwidth]{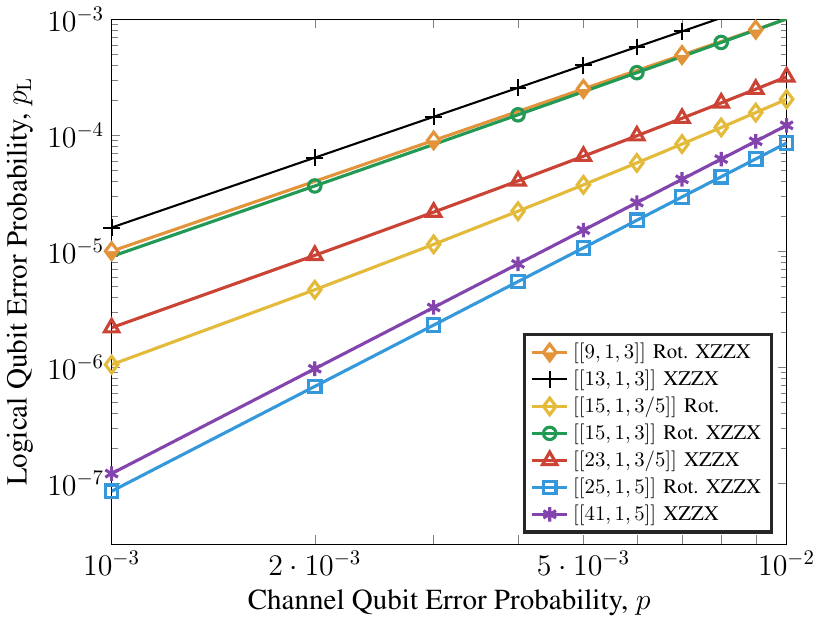}
	%\resizebox{0.99\columnwidth}{!}{
	    %\input{Figures/PerformanceComparison}
	%}
	\caption{Logical error probability vs. physical error probability of the channel. In the plot are reported several topological planar codes of possible interest over an asymmetric channel with $A=10$. ``Rot." stands for rotated surface codes.} 
	\label{Fig:plot_A1/A100}
\end{figure}

In Fig.~\ref{Fig:plot_A1/A100}
we depict some XZZX and surface codes over a channel with asymmetry $A=10$. 
%The plot shows three different groups of curves. 
Among the codes with distance $d=3$, the rotated $[[9,1,3]]$ XZZX code is to be preferred due to its good performance by using the lowest number of qubits.
This shows that the rotation technique (i.e., ad-hoc puncturing of the code) is able to obtain an increase of the coding rate without deteriorating too much the performance. In some cases, the puncturing even improves it.
As expected, the combination of a rectangular lattice, rotation, and XZZX deteriorates the performance, and here it is shown by the fact that the $[[15,1,3]]$ rotated XZZX code (derived from the $[[15,1,3/5]]$ rotated code) achieves the same performance of the shortest planar code (i.e., the $[[9,1,3]]$).
Moving to lower coding rates, well-design rectangular lattices show a higher $\M{Z}$ error correction capability. 
For this reason, they represent a good compromise between performance and codeword length. 
On the other hand, code with large distance $d$ as the $[[25,1,5]]$ and the $[[41,1,5]]$ show the best performance. 
Finally, it is visible in the plot how the code distance influences the slope of the performance curve.
For asymmetric lattices under the asymptotic condition $p \ll 1$ we have that when $A<\infty$ they have the same slope of $d=3$ codes, while for $A=\infty$ the same slope of $d=5$ codes.
%Specifically, the $[[25,1,5]]$ rotated XZZX code has the greatest error correction capability, since it employs less qubits, and hence presents a lower probability for channel errors to occur.  
%

\subsubsection{Effect of channel asymmetry on topological planar codes} 

\begin{figure}[t]
	\centering
 \includegraphics[width=\columnwidth]{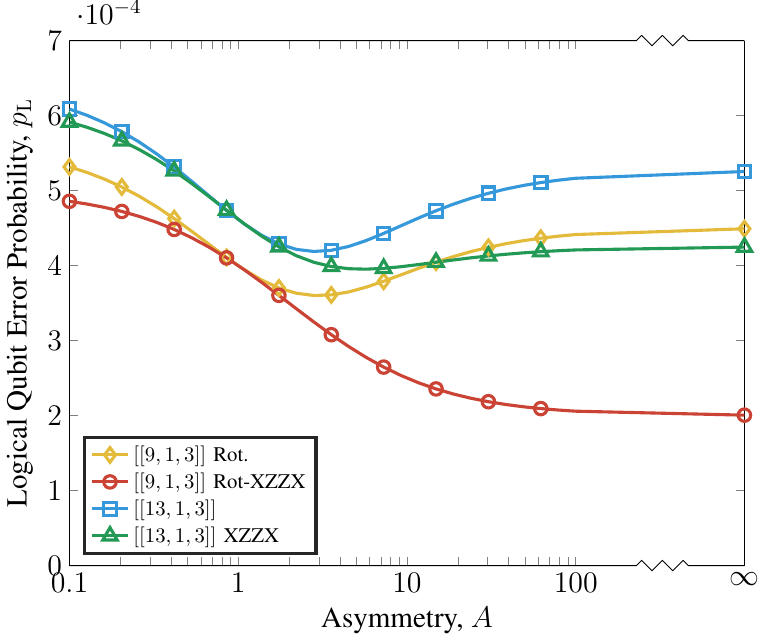}
%	\resizebox{0.99\columnwidth}{!}{
	    %\input{Figures/Asymmetryd3}
	%}
	\caption{Effect of channel asymmetry on the logical error rate,  surface codes with $d=3$, $p=5\cdot 10^{-3}$.  ``Rot." stands for rotated surface codes.}  
	\label{Fig:AsymPlotd3}
\end{figure}

\begin{figure}
	\centering
 \includegraphics[width=\columnwidth]{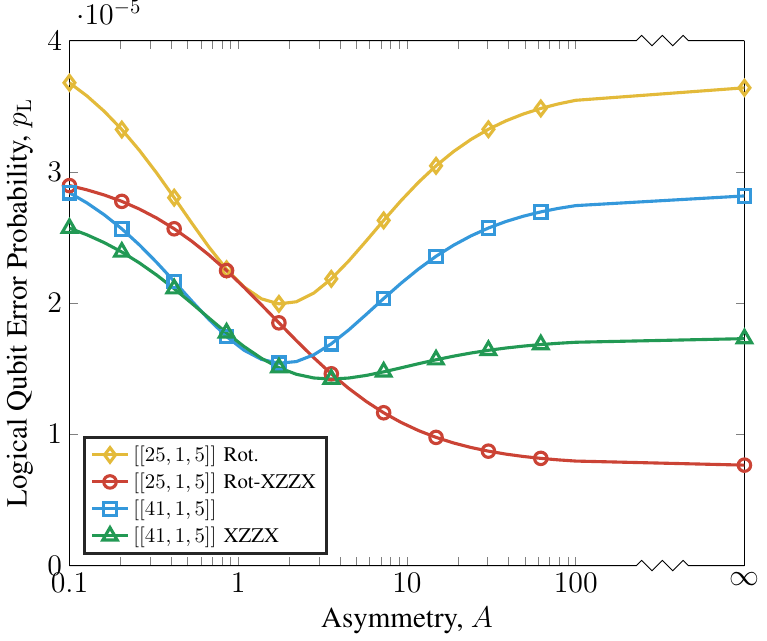}
	%\resizebox{0.99\columnwidth}{!}{
	    %\input{Figures/Asymmetryd5}
	%}
	\caption{Effect of channel asymmetry on the logical error rate,  surface codes with $d=5$, $p=5\cdot 10^{-3}$. ``Rot." stands for rotated surface codes.} 
	\label{Fig:AsymPlotd5}
\end{figure}

\begin{figure}
	\centering
 \includegraphics[width=\columnwidth]{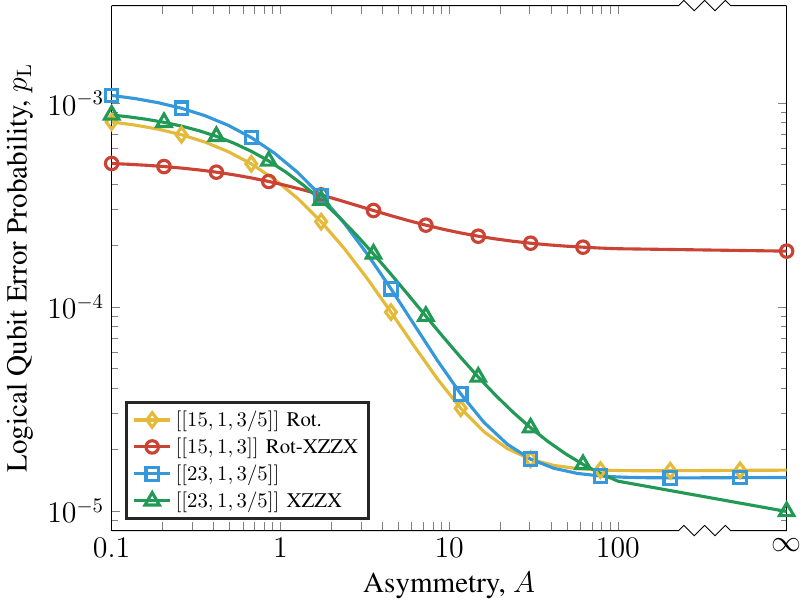}
	%\resizebox{0.99\columnwidth}{!}{
	    %\input{Figures/Asymmetrydx3dz5}
	%}
	\caption{Effect of channel asymmetry on the logical error rate,  surface codes with $d_\mathrm{X}=3/d_\mathrm{Z}=5$, $p=5\cdot 10^{-3}$. ``Rot." stands for rotated surface codes.} 
	\label{Fig:AsymPlotdx3dz5}
\end{figure}

In Fig.~\ref{Fig:AsymPlotd3}, Fig.~\ref{Fig:AsymPlotd5}, and Fig.~\ref{Fig:AsymPlotdx3dz5} we show how the logical error rate is affected by the channel asymmetry $A$, for all considered surface codes. 
To this aim, we fix $p = 5\cdot 10^{-3}$ to be in the $p \ll 1$ regime.

In Fig.~\ref{Fig:AsymPlotd3} we report the logical error rate for topological planar codes with $d=3$.
Since they have the same code distance the relation between them is independent by the chosen $p$.
As expected, we have that XZZX technique does not provide any advantage when $A=1$, while it shows a clear performance improvement in the presence of channel asymmetries.
More than the performance boost itself, it is worth noting that they do not deteriorate when $A$ is increasing. 
This means that we can design the system for the depolarizing channel if we do not know the exact value of $A$, just knowing that $A\geq 1$.
The rotated codes for $d=3$ are able to achieve advantages both in terms of codeword length and performance compared to the non-rotated counterparts. We can conclude that the rotated $[[9,1,3]]$ XZZX code is the best choice among the surface codes with distance $d=3$. 

On the contrary, for $d=5$, although with puncturing we reduce the codeword length, the performance of the rotated codes slightly degrades on the depolarizing channel as shown in Fig.~\ref{Fig:AsymPlotd5}, due to the larger number of error patterns beyond the minimum distance that the decoder can correct. However, given the reduced number of qubits, the small loss for the depolarizing channel, and the performance improvement for asymmetric channels, the rotated $[[25,1,5]]$ XZZX code can be considered the best choice among those with $d=5$.  

Finally, in Fig.~\ref{Fig:AsymPlotdx3dz5} we report codes constructed on rectangular lattices with dimension $3 \times 5$. The surface codes are the $[[23,1,3/5]]$, with and w/o XZZX, and its rotated versions $[[15,1,3/5]]$ and $[[15,1,3]]$ XZZX.  
As highlighted before, the latter is the only one not able to guarantee $d_\mathrm{Z}=5$. 

%

%
% \begin{figure}[t]
% 	\centering
% 	\resizebox{0.45\textwidth}{!}{
% 	    \input{Figures/plotXZZX}
% 	}
% 	\caption{Asymmetry Plot $d=3/5$.}
% 	\label{Fig:AsymPlotdx3dz5}
% \end{figure}
%

%
%\begin{figure}[t]
	%\centering
	%\resizebox{0.45\textwidth}{!}{
	    %\input{Figures/plot_varA_13_XZZX2.tex}
	%}
	%\caption{i rapporti fra curve somo uguali al variaer di ro}
	%\label{Fig:AsymPlotdx3dz5}
%\end{figure}
%

%\newpage

%%%%%%%%%%%%%%%%%%%%%%%%%%%%%%%%%%%%%%%%%%%%%%%%%%%%%%
\section{Conclusions}\label{sec:conclusions}

We have conducted a comprehensive analysis of state-of-the-art topological planar quantum codes, specifically focusing on the XZZX and rotated surface codes. Our primary objective was to derive analytical performance equations for these codes when decoded using complete decoders, like the low-complexity \ac{MWPM}-based decoder. 
Our proposed approach has proven to be highly effective in providing accurate code logical error rates across a wide range of quantum channels, including depolarizing, phase-flip, and asymmetric channels. 
Significantly, our analysis has revealed compelling evidence that examining the error patterns of weight $\lfloor (d-1)/2 \rfloor + 1$ for codes with distance $d$ allows for an accurate description of the coding performance. 
Based on the analysis, we have made noteworthy discoveries regarding specific code variants. For instance, our research demonstrates that the rotated $[[9,1,3]]$ XZZX code outperforms other surface codes with a distance of 3, making it an optimal choice. Similarly, among the codes with a distance of 5, the rotated $[[25,1,5]]$ XZZX code showcases superior performance. Additionally, we have established that the combined use of rotation and XZZX variants can lead to a degradation in performance when starting from a rectangular lattice designed to provide asymmetric error correction capabilities. 
The tables presented in our paper provide a valuable resource for computing the performance of surface codes over arbitrary asymmetric quantum channels. This tool facilitates the analysis and design of complex systems that employ error correction techniques.

%%%%%%%%%%%%%%%%%%%%%%%%%%%%%%%%%%%%%%%%%%%%%%%%%%%%%%

%\section*{Acknowledgment}
%This work has been carried out in the framework of the CNIT National Laboratory WiLab.

\bibliographystyle{IEEEtran}
\bibliography{Files/IEEEabrv,Files/StringDefinitions,Files/StringDefinitions2,Files/refs}
%%%%%%%%%%%%%%%%%%%%%%%%%%%%%%%%%%%%%%%%%%%%%%%%%%%%%%

%\section{Appendix}\label{sec:Appendix}

\end{document}